\documentclass[12pt,a4paper]{article}

\newtheorem{theorem}{Theorem}
\newtheorem{acknowledgement}[theorem]{Acknowledgement}

\input{tcilatex}
\begin{document}

\title{Physical aspects of the field-theoretical description \\
of two-dimensional ideal fluids}
\author{Florin Spineanu and Madalina Vlad \\
National Institute of Laser, Plasma and Radiation Physics \\
Magurele, 077125 Bucharest, Romania}
\maketitle

\begin{abstract}
The two-dimensional ideal (Euler) fluids can be described by the classical
fields of streamfunction, velocity and vorticity and, in an equivalent
manner, by a model of discrete point-like vortices interacting in plane by a
self-generated long-range potential. This latter model can be formalized, in
the continuum limit, as a field theory of scalar matter in interaction with
a gauge field, in the $su\left( 2\right) $ algebra. This description has
already offered the analytical derivation of the \emph{sinh}-Poisson
equation, which was known to govern the stationary coherent structures
reached by the Euler fluid at relaxation. In order this formalism to become
a familiar theoretical instrument it is necessary to have a better
understanding of the physical meaning of the variables and of the operations
used by the field theory. Several problems will be investigated below in
this respect.
\end{abstract}

\section{Introduction}

The two-dimensional incompressible ideal fluid is governed by the Euler
equation%
\begin{equation}
\frac{d\mathbf{\omega }}{dt}=0  \label{eq10}
\end{equation}%
where $\mathbf{\omega }$ is the vorticity $\mathbf{\omega =\nabla \times v}$%
, a vector perpendicular on the plane where the flow with the velocity $%
\mathbf{v}$ is contained. It is useful to define the streamfunction $\psi $
from which the velocity field is derived $\mathbf{v}=\widehat{\mathbf{e}}%
_{z}\times \mathbf{\nabla }\psi $ and the vorticity is $\mathbf{\omega =}%
\widehat{\mathbf{e}}_{z}\Delta \psi $. Most of the results in the theory of
the ideal $2D$ Euler fluid have been obtained using these three quantities,
which have clear physical meaning and are measurable experimentally.

On the other hand the striking result that the asymptotic stationary states
obtained at relaxation exhibit a strong coherency of the flow cannot be
easily described in the framework defined by $\left( \psi ,\mathbf{v},%
\mathbf{\omega }\right) $. The stationary flows from Eq.(\ref{eq10}) obey
the equation $\left[ \left( -\mathbf{\nabla }\psi \times \widehat{\mathbf{e}}%
_{z}\right) \cdot \mathbf{\nabla }\right] \mathbf{\nabla }^{2}\psi =0$ which
has a very large space of possible solutions. Clearly the selection of the
final states is dictated by an additional constraint that may have the form
of a functional extremal condition. Finding a functional defined on the
space of flow configuration, as for example a density of a Lagrangian and an
action functional, has not been possible in the traditional approach.

A model (Hamilton, Kirchhoff, Onsager) which is equivalent with the $2D$
Euler fluid consists of a set of discrete point-like vortices interacting in
plane by a potential which is created by themselves and has long range, 
\emph{i.e.} it is Coulombian%
\begin{eqnarray}
\frac{dx_{i}}{dt} &=&-\frac{\partial }{\partial y}\dsum\limits_{j\neq
i}^{N}\omega _{0}\ln \left( \frac{\left\vert \mathbf{x-x}_{j}\right\vert }{L}%
\right)  \label{eq20} \\
\frac{dy_{i}}{dt} &=&\frac{\partial }{\partial x}\dsum\limits_{j\neq
i}^{N}\omega _{0}\ln \left( \frac{\left\vert \mathbf{x-x}_{j}\right\vert }{L}%
\right)  \nonumber
\end{eqnarray}%
The following observation will have an important consequence later in the
theory. In any further development it is not sufficient to only work with
the system (\ref{eq20}). The system (\ref{eq20}) as it is still needs to
specify which are the \emph{objects} whose positions $\left(
x_{i},y_{i}\right) $ evolve in plane.

The elementary objects can be \emph{charges}. For large $N$ the system can
be treated as a statistical ensemble. In particular it has been shown that
it has negative temperature (S.F. Edwards and J. B. Taylor) which suggests
an intrinsic tendency to self-organization into large structures.

On the other hand the discrete elementary objects can be \emph{vortices }as
actually is requested by the equivalence with the Euler fluid, where the
natural logarithm appears as the propagator inverting the equation relating
the vorticity and the streamfunction, $\Delta \psi =\omega $ (Kraichnan,
Montgomery). Again in a statistical approach, the stationary states
corresponding to the maximum entropy under the conservation of the mass,
momentum and energy have been found to be described by the differential
equation \emph{sinh}-Poisson%
\begin{equation}
\Delta \psi +\sinh \psi =0  \label{eq25}
\end{equation}%
where $\psi $ is the streamfunction.

The problem is how to differentiate between the system consisting of Eqs.(%
\ref{eq20}) \emph{plus} the information that the objects are charges and the
system consisting of Eqs.(\ref{eq20}) \emph{plus} the information that the
objects are vortices. S.F. Edwards and J. B. Taylor observe that the
two-dimensional plasma (\emph{i.e.} charges) and the vortex fluid (\emph{i.e.%
} vortices) are described by the same system of equations. The system of 
\emph{charges} and the system of \emph{vortices} are formally identical if
we replace the charge by the circulation. However for statistical
considerations it is not sufficient. The field theory clearly shows this,
obtaining the distinct results that the system of charges is described by
the Liouville equation and the system of vortices by the \emph{sinh}-Poisson
equation. The field theory describes the charges by an Abelian model and a
vortex fluid by a non-Abelian one.

\bigskip

The \emph{sinh}-Poisson equation has been confirmed by numerical simulations
of relaxation of Euler fluid from states of turbulence to the coherent and
cuasi-stationary asymptotic states (Montgomery).

In consequence of the discussion above we note that the derivation of the 
\emph{sinh}-Poisson equation from purely statistical consideration still
requires some care. In addition the statistical approach could not be
extended to other type of fluids like the $2D$ atmosphere or the plasma in
strong magnetic field. However the fact that the streamfunction $\psi $
verifies the \emph{sinh}-Poisson equation has been confirmed after careful
numerical verification. Any new theory of the Euler fluid will have to be
confronted to this challange, to derive the \emph{sinh}-Poisson equation.

\bigskip

The field theoretical (FT) model for the Euler fluid (Spineanu and Vlad,
2003) is able to provide a purely analytical derivation of the \emph{sinh}%
-Poisson equation. Moreover it can be extended to the more complicated
problem of the $2D$ plasma in strong transversal magnetic field and $2D$
planetary atmosphere. The FT models for fluid, plasma, atmosphere have
clearly shown that they are able to derive new results which are
inaccessible to the traditional approaches based on $\left( \psi ,\mathbf{v},%
\mathbf{\omega }\right) $. However these FT models would be more easily
adopted as an instrument of theoretical investigation if there would be a
physical understanding of the meaning of the field theoretical concepts,
operations, etc. in terms of the more familiar $\left( \psi ,\mathbf{v},%
\mathbf{\omega }\right) $.

\bigskip

\section{The field theoretical representation of the elementary vortex as a
global string}

The definition of the elementary object in the discrete model of point-like
vortices consists of two characteristics:

\begin{enumerate}
\item the elementary object in plane is strictly point-like, there is no
spatial extension.We can represent it however as a line extending in the $z$
direction, \emph{i.e.} perpendicular on the plane

\item the elementary object carry a vorticity \textquotedblleft
content\textquotedblright\ which, although it is not a scalar charge, it is
not introduced as the result of a fluid rotating around the vortex
\end{enumerate}

For the reason resulting from the second characteristic, the convenient
representation of the elementary vortex in field theory is the global
string. This is a time-independent solution of the equation of motion of a
spontaneously broken global $U\left( 1\right) $ Higgs model (Davies and
Shellard). By this representation we dispose of a field theory for the
elementary vortex as a Nambu-Goldstone boson $\alpha \left( x\right) $ with
the complex scalar field%
\begin{equation}
\widehat{\phi }=\eta \exp \left( i\alpha \right)  \label{eq27}
\end{equation}%
where $\eta $ is the vacuum value of $\widehat{\phi }$. Later this model
will prove to be essential in the investigation of the fermionic zero-modes
propagating along the string and strictly confined to it. The axial anomaly
related to these modes interacting with a gauge field will allow us to get
an understanding of the vorticity concentration, a process of high
importance in the tropical cyclone and tornadoes.

For the global string there is no Magnus force acting on the string when it
is moving with velocity $\mathbf{v}$ in the fluid. This is not a problem
since the lows of motion of the point-like objects is given by the purely
kinematic equations of motion and the motion resulting from these equation
is not regarded as a physical interaction with a fluid environment.

However we can not be satisfied with the definition of the elementary vortex
as a global string \emph{plus} the statement that it carries a fixed
vorticity. As shown by Davies and Shellard the global string gets a fixed
vorticity when it is assumed that there is interaction between the string
and a background field which breaks the Lorentz invariance and is equivalent
with inducing rotation of the global string around its axis, or a linear
time variation of the bosonic field $\alpha =\theta +const\times t$ with $%
\theta $ the azimuthal angle. Taking the energy density of the pure
background field $p$ the fixed vorticity carried by a global string in such
background is%
\begin{equation}
\mathbf{\omega }_{0}=\left( 4\pi \eta /\sqrt{p}\right) \widehat{\mathbf{e}}%
_{z}  \label{eq28}
\end{equation}%
This is a mechanism by which we can attach a fixed vorticity to the global
string but it is not less arbitrary than defining the elementary object as
carrying a fixed vorticity $\mathbf{\omega }_{0}$.

\bigskip

\section{The physical meaning of the concepts involved in the field
theoretical model of the Euler fluid}

\subsection{Review of field theoretical model for the Euler fluid}

Jackiw and Pi have developed a field-theoretical for the continuum limit of
a system of charges in plane. The equations of motion can be derived from
the density of a Lagrangian and the theory leads to a differential equation
describing the asymptotic stationary states of the system. This is the
Liouville equation.

We have found that the continuum limit of the discrete model of \emph{%
vortices} is given by a field theory with the Lagrangian that is
characterized by: non-relativistic (Schrodinger), Chern-Simons, $4^{th}$
order scalar field self-interaction. The Lagrangian is%
\begin{eqnarray}
L &=&-\kappa \varepsilon ^{\mu \nu \rho }\mathrm{tr}\left( \left( \partial
_{\mu }A_{\nu }\right) A_{\rho }+\frac{2}{3}A_{\mu }A_{\nu }A_{\rho }\right)
\label{eq30} \\
&&+i\mathrm{tr}\left( \phi ^{\dagger }\left( D_{0}\phi \right) \right) -%
\frac{1}{2}\mathrm{tr}\left( \left( D_{k}\phi \right) ^{\dagger }\left(
D^{k}\phi \right) \right)  \nonumber \\
&&+\frac{1}{4\kappa }\mathrm{tr}\left( \left[ \phi ^{\dagger },\phi \right]
^{2}\right)  \nonumber
\end{eqnarray}%
where $\phi $, $\phi ^{\dagger }$, $A_{\mu }$, $A_{\mu }^{\dagger }$ are $%
SU\left( 2\right) $ elements. The covariant derivatives are 
\begin{equation}
D_{\mu }=\partial _{\mu }+\left[ A_{\mu },\right]  \label{eq31}
\end{equation}%
with the metric $g^{00}=-1$, $g^{11}=g^{22}=1$. The equations of motion are 
\begin{eqnarray}
iD_{0}\phi &=&-\frac{1}{2m}D_{k}D^{k}\phi -\frac{1}{2m\kappa }\left[ \left[
\phi ,\phi ^{\dagger }\right] ,\phi \right]  \label{eq32} \\
-\kappa \varepsilon ^{\mu \nu \rho }F_{\nu \rho } &=&iJ^{\mu }  \nonumber
\end{eqnarray}%
where 
\begin{eqnarray}
J^{0} &=&\left[ \phi ^{\dagger },\phi \right]  \label{eq33} \\
J^{k} &=&-\frac{i}{2m}\left( \left[ \phi ^{\dagger },\left( D^{k}\phi
\right) \right] -\left[ \left( D^{k}\phi \right) ^{\dagger },\phi \right]
\right)  \nonumber
\end{eqnarray}%
which is \emph{covariantly conserved} $D_{\mu }J^{\mu }=0$. The action
functional calculated with this Lagrangian has the fundamental property that
can be written in the Bogomolnyi form, \emph{i.e.} as a sum of squares,
which clearly identifies the absolute extrema in the space of states by
simply taking these square terms to zero. The energy is 
\begin{equation}
E=\frac{1}{2}\mathrm{tr}\left( \left( D_{k}\phi \right) ^{\dagger }\left(
D^{k}\phi \right) \right) -\frac{1}{2\kappa }\mathrm{tr}\left( \left[ \phi
^{\dagger },\phi \right] ^{2}\right)  \label{eq34}
\end{equation}%
The Gauss law is the zero component of the second equation of motion 
\begin{equation}
-2\kappa F_{12}=iJ^{0}  \label{eq35}
\end{equation}%
Introducing the notations $D_{\pm }=D_{x}\pm iD_{y}$ and similarly for other
variables, the first term in Eq.(\ref{eq34}) is 
\begin{equation}
\mathrm{tr}\left( \left( D_{k}\phi \right) ^{\dagger }\left( D^{k}\phi
\right) \right) =\mathrm{tr}\left( \left( D_{-}\phi \right) ^{\dagger
}\left( D_{-}\phi \right) \right) -i\mathrm{tr}\left( \phi ^{\dagger }\left[
F_{12},\phi \right] \right)  \label{eq36}
\end{equation}%
Replacing in the expression of the energy, the last term, coming from the 
\emph{potential} energy or the scalar field self-interaction is canceled and
we obtain 
\begin{equation}
E=\frac{1}{2}\mathrm{tr}\left( \left( D_{-}\phi \right) ^{\dagger }\left(
D_{-}\phi \right) \right)  \label{eq37}
\end{equation}%
which leads to the equation for the states which correspond to the lowest
energy 
\begin{equation}
D_{-}\phi =0  \label{eq38}
\end{equation}

Then the equations of motion are replaced by 
\begin{eqnarray}
D_{-}\phi &=&0  \label{eq190} \\
\partial _{+}A_{-}-\partial _{-}A_{+}+\left[ A_{+},A_{-}\right] &=&\frac{1}{%
\kappa }\left[ \phi ,\phi ^{\dagger }\right]  \nonumber
\end{eqnarray}%
The solutions are stationary flow configurations obeying the set (\ref{eq190}%
) of two first order partial differential equations from which, under a
reasonable algebraic \emph{ansatz} one can derive the \emph{sinh}-Poisson
equation, a unique differential equation for the streamfunction. The
algebraic ansatz uses the generator of the Cartan sub-algebra and the two
ladder generators%
\begin{eqnarray}
\phi &=&\phi _{1}E_{+}+\phi _{2}E_{-}  \label{eq611} \\
\phi ^{\dagger } &=&\phi _{1}^{\ast }E_{-}+\phi _{2}^{\ast }E_{+}  \nonumber
\end{eqnarray}%
\begin{eqnarray}
A_{+} &=&A_{x}+iA_{y}=aH  \label{612} \\
A_{-} &=&A_{x}-iA_{y}=-a^{\ast }H  \nonumber
\end{eqnarray}%
from which the first SD equation leads to%
\begin{equation}
a=\frac{\partial }{\partial z^{\ast }}\ln \left( \phi _{1}^{\ast }\right)
\label{453}
\end{equation}%
\begin{equation}
a^{\ast }=\frac{\partial }{\partial z}\ln \left( \phi _{1}\right)
\label{454}
\end{equation}%
and%
\begin{eqnarray}
\mathrm{tr}\left( \phi \phi ^{\dagger }\right) &=&\rho _{1}+\rho _{2}
\label{1484} \\
\left[ \phi ,\phi ^{\dagger }\right] &=&\left( \rho _{1}-\rho _{2}\right) H 
\nonumber
\end{eqnarray}%
with the notation%
\begin{eqnarray}
\rho _{1} &\equiv &\left\vert \phi _{1}\right\vert ^{2}  \label{49} \\
\rho _{2} &\equiv &\left\vert \phi _{2}\right\vert ^{2}  \nonumber
\end{eqnarray}%
Using the algebraic ansatz in Eq.(\ref{eq190}) we obtain the \emph{sinh}%
-Poisson equation.

The particular property of the system allowing the Bogomolnyi form of the
action is called Self-Duality. It is worth to mention that the field
theoretical formalism for the Euler fluid has been able to show that the
asymptotic coherent structures of the Euler fluid belong to the same family
as all the other structures known: the solitons, the instantons, all are
solutions of equations derived at Self-Duality. It would be desirable to
have the physical meaning of this concept in terms of classical fluid
variables, \emph{i.e.} $\left( \psi ,\mathbf{v},\mathbf{\omega }\right) $.

\bigskip

\subsection{Connections derived from the fermionic nature of the elementary
point-like vortex}

The elementary object of the discrete model is a point-like vortex. The
magnitude is the same for all elementary vortices, say $\omega _{0}$. It is
not admitted to work with multiple vortices as single objects, for example
consisting of two elementary objects superposed into a unique vortex of
magnitude $2\omega _{0}$. This can be expressed by saying that two
elementary vortices cannot be in the same point. There are only two possible
orientations, given by $\pm \omega _{0}$. We conclude that there are
similarities between the elementary point-like vortex and a spin-$1/2$
object. Further this means that if one looks for a field theoretical
formulation of the model of point-like vortices the most natural way would
involve fermionic fields.

On the other hand, Jackiw and Pi have shown that a system of point electric
charges moving in plane according to the same Eqs.(\ref{eq20}) can be
described by a classical Abelian field theory of a bosonic scalar and gauge
fields with Chern-Simons term and with $\varphi ^{4}$ scalar field
self-interaction. This model has self-dual states that are described by the
Liouville equation, $\Delta \psi +\exp \left( \psi \right) =0$. Since
however the Euler fluid is described by point-like \emph{vortices} instead
of \emph{electric charges}, the field theory must be re-formulated to
reflect the spinorial nature of the elementary objects. Since the spinors
are the lowest representations of the Lorentz group whose complex covering
is $SL\left( 2,\mathbf{C}\right) $ we expect to represent the spinorial
nature of the elementary vortices by taking all bosonic variables of the
Jackiw-Pi model as elements of the $sl\left( 2,\mathbf{C}\right) $ algebra.
This is the model of Eq.(\ref{eq30}) from which the \emph{sinh}-Poisson
equation has been derived. This places the non-Abelian, bosonic scalar field
model, with Chern-Simons term for the gauge field and scalar potential
nonlinearity of order four Eq.(\ref{eq30}) as the main framework where we
should look for more conventional physical significance. However we also
note the possibility that neighboring \ \textquotedblleft
fermionic\textquotedblright\ models can be useful. The connection should be
realised via bosonization procedures, where it is possible.

There are several two-dimensional models that present similarities with the
model of point-like vortices and that have received considerable attention,
with various purposes: the Thirring model, the Schwinger model, the
Nambu-Jona-Lasinio model, etc.

We note that the scalar-field self-interaction of order four, which is in
Non-Abelian form%
\begin{equation}
\frac{1}{2}\mathrm{tr}\left( \left[ \phi ^{\dagger },\phi \right] ^{2}\right)
\label{eq200}
\end{equation}%
or, in Abelian case 
\begin{equation}
\frac{1}{2}\left( \phi ^{\ast }\phi \right) ^{2}  \label{eq201}
\end{equation}%
is the same as the fourth order term in the expansion%
\begin{equation}
\cos \phi -1\approx -\frac{1}{2}\left( \phi ^{\ast }\phi \right) +\frac{1}{24%
}\left( \phi ^{\ast }\phi \right) ^{2}  \label{q202}
\end{equation}

The term $\left( \phi ^{\ast }\phi \right) $ should be considered separately
together with the mass term. If the fourth order scalar-field non-linearity $%
\sim \left( \phi ^{\ast }\phi \right) ^{2}$ comes from $\cos \left( \phi
\right) -1$ then the nonlinearity is the same as in the \emph{sine-Gordon}
model. Then the fermionic model to which we should look is the Thirring
model with the Lagrangian%
\begin{eqnarray}
\mathcal{L}_{Th} &=&-\overline{\psi }\left( i\NEG{\partial}_{\mu }\right)
\psi  \label{q203} \\
&&-\frac{1}{2}\left( \overline{\psi }\gamma ^{\mu }\psi \right) \left( 
\overline{\psi }\gamma _{\mu }\psi \right)  \nonumber
\end{eqnarray}%
which is characterized by a current-current ($JJ$) interaction. Writting the
nonlinearity as%
\begin{equation}
\left( \overline{\psi }\gamma ^{\mu }\psi \right) \left( \overline{\psi }%
\gamma _{\mu }\psi \right) =\frac{1}{2}\left[ \left( \overline{\psi }\psi
\right) ^{2}-\left( \overline{\psi }\gamma _{5}\psi \right) ^{2}\right]
\label{q204}
\end{equation}

We propose the following identification%
\begin{eqnarray}
\left( \overline{\psi }\psi \right) ^{2} &=&\left( \rho _{1}+\rho
_{2}\right) ^{2}  \label{eq205} \\
\left( \overline{\psi }\gamma _{5}\psi \right) ^{2} &=&\left( \rho _{1}-\rho
_{2}\right) ^{2}  \nonumber
\end{eqnarray}%
such that%
\begin{eqnarray}
&&\frac{1}{2}\left[ \left( \overline{\psi }\psi \right) ^{2}-\left( 
\overline{\psi }\gamma _{5}\psi \right) ^{2}\right]  \label{eq206} \\
&=&\frac{1}{2}\left[ \left( \rho _{1}+\rho _{2}\right) ^{2}-\left( \rho
_{1}-\rho _{2}\right) ^{2}\right]  \nonumber \\
&=&2\rho _{1}\rho _{2}  \nonumber
\end{eqnarray}%
By this identification $JJ$ is connected with the product%
\begin{equation}
\rho _{1}\rho _{2}  \label{eq207}
\end{equation}%
which plays a major role in the field theory for Euler. The extremum of the
action for the Thirring model is%
\begin{equation}
JJ=\text{const}  \label{eq208}
\end{equation}%
and this means, after normalizations,%
\begin{equation}
\rho _{1}\rho _{2}=1  \label{eq209}
\end{equation}

This equation plays a major role in the field-theoretical derivation of 
\emph{sinh}-Poisson equation for Euler and demands an interpretation in
physical terms. The identification proposed above can open the way to a
physical interpretation.

The two components are%
\begin{eqnarray}
&&\left( \overline{\psi }\psi \right) =\psi ^{\dagger }\gamma ^{0}\psi =\psi
^{\dagger }\sigma ^{3}\psi \ \ \text{the density of spin}\ \   \label{eq210}
\\
&&\left( \overline{\psi }\gamma _{5}\psi \right) \ \ \text{chiral current} 
\nonumber
\end{eqnarray}

We recall (\textbf{Coleman, Faber and Ivanov}) that the two models: \textbf{%
sine-Gordon} (bosonic) and \textbf{Thirring} (fermionic) are equivalent if $%
4\pi /\beta ^{2}=1+g/\pi $ and the fermionic field $\psi \left( x\right) $
respectively the bosonic field $\theta \left( x\right) $ satisfy the Abelian
bosonization relation%
\begin{equation}
m\overline{\psi }\left( x\right) \left( \frac{1\mp \gamma ^{5}}{2}\right)
\psi \left( x\right) =-\frac{\alpha }{\beta ^{2}}\exp \left[ \pm i\beta
\theta \left( x\right) \right]  \label{eq500}
\end{equation}%
(here the constants $g$, $\beta $, $\alpha $, $m$, are constants). \textbf{%
Witten} uses the notations%
\begin{equation}
\overline{\psi }\left( x\right) \left( \frac{1\pm \gamma ^{5}}{2}\right)
\psi \left( x\right) \equiv \mathcal{O}_{\pm }  \label{eq211}
\end{equation}%
which are called \emph{chiral densities}. According to the identification
our $\rho _{1,2}$%
\begin{equation}
\rho _{1},\rho _{2}\sim \mathcal{O}_{\pm }  \label{eq212}
\end{equation}%
the coefficients of the ladder generators. Then we should compare%
\begin{eqnarray}
\overline{\psi }\left( x\right) \left( \frac{1+\gamma ^{5}}{2}\right) \psi
\left( x\right) &\Longleftrightarrow &\phi _{1}^{\ast }\phi _{1}\ \ \left( 
\text{from}\ \ \phi _{1}E_{+}\right)  \label{eq213} \\
\overline{\psi }\left( x\right) \left( \frac{1-\gamma ^{5}}{2}\right) \psi
\left( x\right) &\Longleftrightarrow &\phi _{2}^{\ast }\phi _{2}\ \ \left( 
\text{from}\ \ \phi _{2}E_{-}\right)  \nonumber
\end{eqnarray}%
We have%
\begin{eqnarray}
\mathrm{tr}\left( \phi \phi ^{\dagger }\right) &=&\rho _{1}+\rho _{2}
\label{eq214} \\
\left[ \phi ,\phi ^{\dagger }\right] &=&\left( \rho _{1}-\rho _{2}\right) H 
\nonumber
\end{eqnarray}%
\begin{eqnarray}
\rho _{1}-\rho _{2} &=&\left[ \phi ^{\dagger },\phi \right] /H=\text{%
vorticity Euler}\ \omega  \label{eq215} \\
&\downarrow &  \nonumber \\
&&\overline{\psi }\left( x\right) \gamma ^{5}\psi \ \ \left( \text{axial
current}\right)  \nonumber
\end{eqnarray}%
\begin{eqnarray}
\rho _{1}+\rho _{2} &=&\mathrm{tr}\left( \phi \phi ^{\dagger }\right)
\label{eq216} \\
&\downarrow &  \nonumber \\
&&\overline{\psi }\left( x\right) \psi \left( x\right) \ \ \text{density of
spin}  \nonumber
\end{eqnarray}

In conclusion a possible identification is%
\begin{eqnarray}
\mathcal{O}_{+} &=&\overline{\psi }\left( \frac{1+\gamma ^{5}}{2}\right)
\psi =\phi _{1}^{\ast }\phi _{1}=\rho _{1}=\ \ \text{density of
positive-chirality modes}  \label{eq217} \\
\mathcal{O}_{-} &=&\overline{\psi }\left( \frac{1-\gamma ^{5}}{2}\right)
\psi =\phi _{2}^{\ast }\phi _{2}=\rho _{2}=\ \ \text{density of
negative-chirality modes}  \nonumber
\end{eqnarray}%
\begin{eqnarray}
\overline{\psi }\psi &=&\rho _{1}+\rho _{2}=\mathrm{tr}\left( \phi ^{\dagger
}\phi \right) \ \text{density of spin (Boyanovsky)}  \label{eq218} \\
\overline{\psi }\gamma ^{5}\psi &=&\rho _{1}-\rho _{2}=\left[ \phi ^{\dagger
},\phi \right] /H=\omega \ \ \text{axial current}  \nonumber
\end{eqnarray}

From Eq.(\ref{eq500}) we have%
\begin{eqnarray}
\overline{\psi }\psi -\overline{\psi }\gamma ^{5}\psi &=&-\frac{2\alpha }{%
m\beta ^{2}}\exp \left( i\beta \theta \right)  \label{eq219} \\
\overline{\psi }\psi +\overline{\psi }\gamma ^{5}\psi &=&-\frac{2\alpha }{%
m\beta ^{2}}\exp \left( -i\beta \theta \right)  \nonumber
\end{eqnarray}%
\begin{eqnarray}
\overline{\psi }\psi &=&-\frac{2\alpha }{m\beta ^{2}}\cos \left( \beta
\theta \right)  \label{eq220} \\
\overline{\psi }\gamma ^{5}\psi &=&i\frac{2\alpha }{m\beta ^{2}}\sin \left(
\beta \theta \right)  \nonumber
\end{eqnarray}%
\begin{eqnarray}
\left( \overline{\psi }\psi \right) ^{2}-\left( \overline{\psi }\gamma
^{5}\psi \right) ^{2} &=&\left( \frac{2\alpha }{m\beta ^{2}}\right) ^{2}%
\left[ \left( \cos \left( \beta \theta \right) \right) ^{2}+\left( \sin
\left( \beta \theta \right) \right) ^{2}\right]  \label{eq221} \\
&=&\left( \frac{2\alpha }{m\beta ^{2}}\right) ^{2}  \nonumber
\end{eqnarray}%
which could correspond to the equation%
\begin{equation}
\left( \rho _{1}+\rho _{2}\right) ^{2}-\left( \rho _{1}-\rho _{2}\right)
^{2}=\left( \frac{2\alpha }{m\beta ^{2}}\right) ^{2}  \label{eq222}
\end{equation}%
or%
\begin{equation}
\rho _{1}\rho _{2}=\frac{1}{4}\left( \frac{2\alpha }{m\beta ^{2}}\right) ^{2}
\label{eq223}
\end{equation}%
The fact that none of $\rho _{1}$ or $\rho _{2}$ can be zero results from
the fact that each represents a \emph{chiral density}%
\begin{eqnarray}
\mathcal{O}_{+} &=&\rho _{1}  \label{eq224} \\
\mathcal{O}_{-} &=&\rho _{2}  \nonumber
\end{eqnarray}

The vanishing of any of the two densities $\rho _{1}$ or $\rho _{2}$ would
lead to the equation%
\begin{equation}
\exp \left( i\theta \right) =0  \label{eq225}
\end{equation}%
with no solution. The fact that none of these two functions can vanish is
essential for the structure of the differential equation describing the
self-dual states: the \emph{sinh}-Poisson equation contains both $\exp
\left( +\psi \right) $ and $\exp \left( -\psi \right) $. If we think to the
derivation of this equation, reviewed above, it leaves the impression that
the value of the vorticity in a particular point is always a result of two
opposite actions, and none of them can be absent: creation of vorticity in a
small spatial patch by the effect of densification of point-like vortices by
the action of the ladder generator $E_{+}$ followed by decrease of the local
vorticity by rarefaction of elementary vortices in the same patch, realised
by the second generator $E_{-}$. Actually we see that the FT operations mean
the substraction of the two chiral densities $O_{+}-O_{-}$ such that the
spin density is eliminated leaving only the chiral current, \emph{i.e.}
vorticity.

\bigskip

\subsection{Review of the connections and comments on the interpretation}

The connection goes through these steps:

\begin{itemize}
\item the Euler equation

\item the point-like vortices

\item the Jackiw-Pi model, constructed for \emph{charged} point-like objects

\item the non-relativistic Non-Abelian $SU\left( 2\right) $ CS $4^{th}$;
this is for point-like \emph{vortices}.

\item observation that the scalar-field part is close at low amplitudes to
the sine-Gordon model with the nonlinearity $\cos \phi -1$ expanded to
second order, \emph{i.e.} to $\phi ^{4}$ term.

\item exploiting by \textquotedblleft anti-bosonization\textquotedblright\
the connection with theThirring \emph{fermionic} non-linearity, $JJ$.

\item identification of the \emph{chiral densities} defined within the
Thirring model as the amplitudes of the algebraic operations of the ladder
generators in the \emph{ansatz} for the $su\left( 2\right) $ scalar field $%
\phi $ of the FT model for Euler.

\item $JJ$ constant, leading to SD states
\end{itemize}

\bigskip

If this identifications are confirmed then it formulates the field
theoretical model in a more accessible way, preparing for physical
discussions in terms of classical fluid approaches.

We need a non-Abelian field theory (FT) and TWO functions, $\rho _{1}$ and $%
\rho _{2}$ to describe the Euler fluid. This comes from the nature of the
point-like object, which is a \emph{vortex} is not a charge like in
Jackiw-Pi. For Jackiw-Pi model an Abelian charge was sufficient and the
result was the Liouville equation. The intermediate Thirring-sine-Gordon
mapping (bosonization) is fully \emph{Abelian}. Since for the Euler fluid
the point-like objects carry vorticity the model must be \emph{non-Abelian}
and the final equation at self-duality is \emph{sinh}-Poisson.

Taking the physical streamfunction as $\Psi $, at self-duality in FT we have%
\begin{eqnarray}
\left( \overline{\psi }\gamma _{5}\psi \right) &=&\omega  \label{eq301} \\
&=&\rho _{1}-\rho _{2}  \nonumber \\
&=&\sinh \Psi  \nonumber
\end{eqnarray}%
and%
\begin{eqnarray}
\left( \overline{\psi }\psi \right) &=&\rho _{1}+\rho _{2}  \label{eq302} \\
&=&\cosh \Psi  \nonumber
\end{eqnarray}%
and 
\begin{eqnarray}
JJ &=&\left( \overline{\psi }\psi \right) ^{2}-\left( \overline{\psi }\gamma
_{5}\psi \right) ^{2}  \label{eq303} \\
&=&\left( \cosh \Psi \right) ^{2}-\left( \sinh \Psi \right) ^{2}  \nonumber
\\
&=&1  \nonumber
\end{eqnarray}%
which means that SD is equivalent with constant interaction $JJ$.

\bigskip

\section{Discussion of the meaning of $\protect\rho _{1}\protect\rho _{2}=1$
in the non-Abelian model}

We have seen the meaning of the relation $\rho _{1}\rho _{2}=1$ in the case
where we use the fermionic counterpart of the theory. Now we return to the
bosonic, non-Abelian model.

For the Euler fluid the condition $\rho _{1}\rho _{2}=1$ arises from the
equality of two distinct expression for the magnetic field $F_{+-}$
calculated first with only $\phi _{1}$ and alternatively with only $\phi
_{2} $. This is possible because the algebraic operation involved in the
first self-duality equation 
\begin{equation}
D_{-}\phi =0  \label{eq304}
\end{equation}%
preserves without mixing the generators of the algebra in the original
places where they exist according to the algebraic \emph{ansatz}. The
commutators reproduce the same generator (with change of sign for $E_{-}$)
and the equality with zero gives two equation in which the potentials $%
A_{\pm }$ are expressed either by $\phi _{1}$ or by $\phi _{2}$. It is now
essential that the algebraic ansatz connects formally the two potentials $%
A_{+}=aH$ and $A_{-}=-a^{\ast }H$. The operations that are performed in the
equation $D_{-}\phi =0$ involve commutators of $A_{-}$ (from the covariant
operator $D_{-}$) with the function $\phi $, or%
\begin{equation}
\left[ H,E_{\pm }\right] =\pm 2E_{\pm }  \label{eq327}
\end{equation}%
It is essential that the commutators do NOT mix the algebra generators. If $%
A_{-}$ were \emph{not} along the Cartan generator $H$ but it contained $%
E_{\pm }$ then there would have been mixing of generators.They remain
separate and the two equations derived from $D_{-}\phi =0$ involve $\phi
_{1} $ and respectively $\phi _{2}$, separately. This looks like an equation
of eigenfunctions where the generator $H$ (which is the potential~$A_{-}$)
leaves invariant the ladder $E_{+}$, as if it was its eigenfunction. This
means that the potential which is the fluid physical velocity does not
affect the ladder generator when it is compatible with it.

It is equally important that the function $a$ and its complex conjugate $%
a^{\ast }$ are the potentials $A_{\pm }$. This means physically to consider
the same velocity in the transversal plane, acting to determine $\phi _{1}$
and respectively $\phi _{2}$. The two functions $A_{+}$ and $A_{-}$
represent, in this algebraic \emph{ansatz} the same physical velocity in the
plane but seen from two opposite directions, from up and from down but along
the same $z$ direction. This means that $\phi _{1}$ and $\phi _{2}$ must be
connected and finally the relationship 
\begin{equation}
\rho _{1}\rho _{2}=1  \label{eq337}
\end{equation}%
is obtained.

We can express the magnetic field $F_{+-}$ in two ways, using each time only 
\emph{one} variable $\phi _{1}$ or $\phi _{2}$.

In conclusion, we have two ways to think the Eq.(\ref{eq337}): $JJ=$const,
in every point. The other way is:%
\[
\Delta \ln \left\vert \phi _{1}\right\vert ^{2}+\Delta \ln \left\vert \phi
_{2}\right\vert ^{2}=0 
\]%
coming from the fact that we can express the field $F_{+-}$ in two different
ways, using either $\phi _{1}$ (which means increase by $E_{+}$) or $\phi
_{2}$ (using $E_{-}$).

This is the reason for $\rho _{1}\rho _{2}=1$.

\section{Conclusion}

Less than a physical interpretation, the present discussion may be useful
for further investigation of the connection between the traditional approach
to the fluid physics and the field theoretical approach.

\begin{acknowledgement}
This work has been supported by the Romanian Ministry of Education and
Research under Ideas Exploratory Research Project No.557
\end{acknowledgement}

\end{document}